\documentclass[preprint,aps,nofootinbib]{revtex4}
\usepackage{epsfig}
\usepackage{bm}

\def\vtau{{\vec \tau}}
\def\hpiNN{{h^1_\pi}}

\def\mbfJ{{\bm J}}
\def\mbfk{{\bm k}}

\def\mbfp{{\bm p}}

\def\mbfr{{\bm r}}

\def\vsig{{\bm \sigma}}
\def\vtau{{\bm \tau}}

\def\vto{\tilde{v}^{3p1}_{t(d)}(r)}

\begin{document}
%\hfill{\tt pncAP4}

\title{Parity nonconserving observables in thermal neutron
capture on a proton}
\author{
C. H. Hyun$^{a,\, b}$\footnote{e-mail: hch@meson.skku.ac.kr},
S. J. Lee$^{a}$,
J. Haidenbauer$^{c}$,
S. W. Hong$^{a}$}

\affiliation{
(a) Department of Physics and
Institute of Basic Science, \\
Sungkyunkwan University, Suwon 440-746, Korea\\
(b) School of Physics, Seoul Nat'l University,
Seoul 151-742, Korea \\
(c) Forschugszentrum J\"{u}lich, Institut f\"{u}r
Kernphysik, D-52425 J\"{u}lich, Germany}

\date{November 26, 2004}

\begin{abstract}
We calculate parity nonconserving observables in the
processes where a neutron is captured on a proton 
at the threshold energy radiating a photon.
Various potential models such as Paris, Bonn and Argonne $v18$
are used for the strong interactions, 
and the meson-exchange description is employed for 
the weak interactions between hadrons.
The photon polarization $P_\gamma$ in the unpolarized
neutron capture process and photon asymmetry $A_\gamma$ in the 
polarized neutron capture process are obtained in terms of the 
weak meson-nucleon coupling constants. $A_\gamma$ turns out to 
be basically insensitive to the employed strong interaction models
and thus can be uniquely determined in terms of the weak coupling constants,
but $P_\gamma$ depends significantly on the strong interaction models.
\end{abstract}
\maketitle
\setcounter{footnote}{0}

\section{Introduction}

Recent experiments to explore the weak interactions between
hadrons through parity nonconserving (PNC) observables in 
nuclear systems \cite{cs133,bwprl-99} or 
reactions \cite{pp,np} have triggered revived interests in this field. 
These PNC observables can be related
to the meson-nucleon weak coupling constants
which are introduced in the meson-exchange potential description
of the hadronic weak interaction \cite{ddh80}.
However, due to various uncertainties (see Ref. \cite{ddh80} for details),
the weak coupling constants were fixed only within
certain ranges \cite{ddh80}.
Thus, it has been hoped that the PNC observables from various experiments
can reduce the range of these coupling constants and
eventually determine the values.
The situation, however, has not been much improved even by the 
recent measurements. 
For example, the value of the $\pi-N$ weak coupling constant, 
$h^1_\pi$, from the anapole moment of $^{133}$Cs \cite{cs133} is 
inconsistent with a previous value obtained from the 
forbidden $\gamma$-decay of $^{18}$F \cite{ah85}:
$h^1_\pi$ determined by the former is larger than that from the 
latter by a factor of 7. 
New experiments, already completed
\cite{pp}, being done \cite{np} or expected to be performed,
concern two-nucleon systems in which many body effects are absent. 
Thus they are expected to give more stringent constraints on the 
weak coupling constants.
For the current status of research on the weak coupling constants,
see \cite{des-physrep}.

In this work, we calculate the photon asymmetry $A_\gamma$ 
in the radiative capture of a polarized neutron on a proton,
$\vec{n} + p \rightarrow d + \gamma$, and the circular
polarization of photons $P_\gamma$ in $n + p \rightarrow 
d + \gamma$ at the threshold. 
The latest experimental value of
$A_\gamma$ is $- (1.5 \pm 4.8) \times 10^{-8}$ \cite{ill88},
but the experiment being done at LANSCE aims at the accuracy
of $10^{-9}$ \cite{np}. 
Theoretical calculations of $A_\gamma$ using 
strong models made in the 1960's and the 1970's 
such as Hamada-Johnston, Reid-soft-core and Tourreil-Sprung
show results similar to each other; 
$A_\gamma \simeq -0.11 h^1_\pi$ \cite{des75}. 
$A_\gamma$ is predominantly determined by $h^1_\pi$ 
and depends very little on other coupling constants 
(as will be shown in Table~\ref{tab:result}). 
In this work we present $A_\gamma$ calculated with 
potentials such as Paris \cite{paris}, Bonn \cite{bonn87}, Bonn-A and 
Bonn-B \cite{bonn-ab}, and Argonne $v18$ (A$v18$)
\cite{av18}. 
We compare our results with previous ones \cite{des75}
and investigate the model dependence of $A_\gamma$.

Contrary to $A_\gamma$, $P_\gamma$ at the threshold is 
known to be sensitive to the heavy meson ($\rho$ and $\omega$)
components of the weak potentials \cite{des75,cra76}.
The most recent experimental value of $P_\gamma$ is 
$(1.8 \pm 1.8) \times 10^{-7}$ \cite{pgam}, 
and theoretical calculations made in the 1970's
agree with this value within the experimental errors. 
However, since $P_\gamma$ is 
sensitive to the short range properties of the strong interactions 
as well as of the weak interactions, 
its model dependence is more noticeable
than $A_\gamma$ \cite{cra76,mck78}. 
Since the inverse process,
$\vec{\gamma} + d \rightarrow n + p$, whose PNC asymmetry at the 
threshold is equal to $P_\gamma$, becomes experimentally feasible 
nowadays, we expect that $P_\gamma$ can be measured 
more precisely and can provide more constraints on the weak 
dynamics of hadrons. We thus investigate the model dependence of 
$P_\gamma$ with the same potentials that we use
in calculating $A_\gamma$.

In Sect. 2, we present the Desplanques-Donaghue-Holstein (DDH) potential
\cite{ddh80} and the parity-admixed wave functions in the initial and the final
states. In Sect. 3, the electromagnetic operators are presented, matrix
elements are derived, and the results for $A_\gamma$ and $P_\gamma$
are shown. 
Discussions on the results follow in Sect. 4.

\section{Parity Admixed Wave Function}

The Schr\"{o}dinger equation for a two-nucleon system can be written as
\begin{eqnarray}
H\, \Psi(\mbfr) &=& \left[ - \frac{1}{m_N} \left(
\frac{1}{r}\frac{\partial^2}{\partial\, r^2}\, r -
\frac{l(l+ 1)}{r^2} \right) + V_C(r) + V_T(r)\, S_{12}(\hat{\mbfr})
+ V_{pnc}(\mbfr) \right]\, \Psi(\mbfr) \nonumber \\
&=& E\, \Psi(\mbfr),
\label{eq:schro}
\end{eqnarray}
where $V_T$ represents the tensor potential
and $V_C$ includes central, spin-orbit, spin-spin and quadratic spin-orbit 
interactions in the strong potential. 
In the Paris and Bonn potentials,
it is essential to include the momentum dependent term
in the central potential
to obtain the correct phase shifts even at low energies. 
In Ref. \cite{pmetalnpa81} 
a transformation useful for treating the momentum dependent
term is suggested. 
In this work, however, we have dealt with the momentum dependent terms 
without using such a transformation and have solved the 
Schr\"{o}dinger equation as it is.
We have confirmed that the solutions thus obtained reproduce the results 
of each potential model \cite{paris,bonn87,bonn-ab}
fairly well with differences less than 1\%.
The small differences 
can be attributed to the use of slightly different values of physical 
quantities in the calculations.
$V_{pnc}$ is the PNC potential, and we use the one given by 
DDH \cite{ddh80}
\begin{eqnarray}
V_{pnc}(\mbfr) &=& V^\pi_{pnc}(\mbfr) + V^\rho_{pnc}(\mbfr)
+ V^\omega_{pnc}(\mbfr), \nonumber \\
V^\pi_{pnc}(\mbfr) &=&
i \frac{g_{\pi NN} \hpiNN}{2\sqrt{2} m_N}
\left(\vtau_1 \times \vtau_2\right)^z
\left(\vsig_1 + \vsig_2\right) \cdot
\left[ \mbfp,\, f_\pi(r) \right], \\
V^\rho_{pnc}(\mbfr) &=&
- \frac{g_{\rho NN}}{m_N} \left[
\left( h^0_\rho\,  \vtau_1 \cdot \vtau_2
+ \frac{1}{2} h^1_\rho\,  (\tau^z_1 + \tau^z_2)
+ \frac{h^2_\rho}{2 \sqrt{6}}\,
(3 \tau^z_1\tau^z_2 - \vtau_1  \cdot \vtau_2)
\right) \times \right. \nonumber \\
& & \Big( (\vsig_1 - \vsig_2)\cdot
\left\{ \mbfp,\, f_\rho(r) \right\}\,
+ i(1 + \chi_\rho)\, (\vsig_1\times \vsig_2)
\, \cdot \left[ \mbfp,\, f_\rho(r) \right] \Big)
\nonumber \\
& + & \left.
i  \frac{h^{1'}_\rho}{2} \left(\vtau_1 \times \vtau_2\right)^z
\left(\vsig_1 + \vsig_2\right) \cdot
\left[ \mbfp,\, f_\rho(r)
\right]
- \frac{ h^1_\rho}{2}\, (\tau^z_1 - \tau^z_2)
(\vsig_1 + \vsig_2)\cdot
\left\{\mbfp,\, f_\rho(r) \right\} \right], %\nonumber \\
\\
V^\omega_{pnc}(\mbfr) &=& -
\frac{g_{\omega NN}}{m_N}
\left[ \left(h^0_\omega + \frac{1}{2}\, h^1_\omega\,
(\tau^z_1 + \tau^z_2) \right) \times \right. \nonumber \\
& & \Big((\vsig_1 - \vsig_2)\cdot
\left\{\mbfp,\, f_\omega(r) \right\}
+ i (1 + \chi_\omega)\, (\vsig_1\times \vsig_2) \cdot
\left[\mbfp,\, f_\omega(r)\right]
\Big) \nonumber \\
& & \left. + \frac{h^1_\omega}{2}\, (\tau^z_1 - \tau^z_2)
(\vsig_1 + \vsig_2)\cdot
\left\{\mbfp,\, f_\omega(r)\right\}
\right],
\label{eq:pncpotential}
\end{eqnarray}
where the strong coupling constants are $g_{\pi NN} = 13.45$, 
$g_{\rho NN} = 2.79$, $g_{\omega NN} = 8.37$ and the anomalous magnetic
moments are $\chi_\rho = 3.71$ and $\chi_\omega = - 0.12$.
The Yukawa functions $f_M(r)$ are defined as
\[
f_M(r) =\frac{{\rm e}^{- m_M r}}{4 \pi r},\ \ \
(M = \pi,\ \rho,\ \omega).
\]
The quantities $h^{\Delta I}_{M}$ represent the weak 
meson-nucleon coupling constants
where $\Delta I$ denotes the isospin transfer.

At the threshold energy, the initial scattering state, 
$n + p$, is dominated by 
the lowest angular momentum state, i.e., the $^1 S_0$ channel,
and higher angular momentum states are suppressed.
Thus in this work we just include the next low-lying state, 
the $^3 S_1 - {}^3 D_1$ partial waves,
where the $^3 D_1$ state is induced by the tensor 
interaction in the initial scattering state.
% and estimate its contribution.
Then the parity-even state of the initial wave function consists of 
the $^1 S_0$, $^3 S_1$ and $^3 D_1$ states.

Since $V_{pnc}$ is a parity-odd operator, it creates opposite parity
components in the wave function. For example, when $V_{pnc}$ is operated
on the $^1 S_0$ state, the isoscalar and isotensor terms of $V_{pnc}$ generate
a $^3 \tilde{P}_0$ admixture, where the tilde denotes the 
parity-admixed components generated from the DDH potential.
Similarly, $^3 \tilde{P}_1$ and $^1 \tilde{P}_1$
admixtures arise from applying the isovector and 
isoscalar components of $V_{pnc}$ to the $^3 S_1 - {}^3 D_1$ state, 
respectively.
The total wave function of the initial state with its parity admixture
at the threshold can be written as
\begin{eqnarray}
\Psi_i (\mbfr) &=& \Psi^{pc}_i(\mbfr) + \Psi^{pnc}_i(\mbfr), \nonumber \\
\Psi^{pc}_i(\mbfr) &=& \frac{1}{\sqrt{4 \pi} r}
\left[ u_s(r)\, \chi_{00}\, \zeta_{10} + 
\left( u_t(r) + \frac{S_{12}(\hat{\mbfr})}{\sqrt{8}} w_t(r)\right)
\chi_{1 S_z}\, \zeta_{00} \right], \label{eq:pci}\\
\Psi^{pnc}_i(\mbfr) &=& - \frac{i}{\sqrt{4 \pi}r} \left[
\sqrt{\frac{3}{8}}\,  \tilde{v}^{3p1}_t(r) (\vsig_1 +\vsig_2)
\, \chi_{1 S_z}\, \zeta_{10}
+ \frac{1}{2}\,  \tilde{v}^{3p0}_t(r) (\vsig_1 -\vsig_2)
\, \chi_{00}\, \zeta_{10} \right] \cdot \hat{\mbfr}, \label{eq:pnci}
\end{eqnarray} 
where $\chi_{S\, S_z}$ and $\zeta_{T\, T_z}$ represent the spin and 
isospin part, respectively.
$u_s$ is the radial part of the wave function for the $^1 S_0$ channel,
$u_t$ for $^3 S_1$ and $w_t$ for $^3 D_1$.
The final state wave function can be written in a similar way as
\begin{eqnarray}
\Psi_f (\mbfr) &=& \Psi^{pc}_f(\mbfr) + \Psi^{pnc}_f(\mbfr), \nonumber \\
\Psi^{pc}_f(\mbfr) &=& \frac{1}{\sqrt{4 \pi} r} 
\left( u_d(r) + \frac{S_{12}(\hat{\mbfr})}{\sqrt{8}} w_d(r)\right)
\chi_{1 S_z}\, \zeta_{00}, \label{eq:pcf} \\
\Psi^{pnc}_f(\mbfr) &=& \frac{i}{\sqrt{4 \pi} r}
\left[\frac{\sqrt{3}}{2}\, \tilde{v}^{1p1}_d(r)(\vsig_1 -\vsig_2)
\, \chi_{1 S_z}\, \zeta_{00} 
- \sqrt{\frac{3}{8}}\,  \tilde{v}^{3p1}_d(r) (\vsig_1 +\vsig_2)
\, \chi_{1 S_z}\, \zeta_{10} \right] \cdot \hat{\mbfr},\label{eq:pncf}
\end{eqnarray}
where $u_d(r)\ (w_d(r))$ is the radial wave function
for the $^3 S_1\ ({}^3 D_1)$ deuteron state, 
and $\tilde{v}^{1p1}_d$ and $\tilde{v}^{3p1}_d$ denote the parity 
nonconserving admixture due to the $^1 \tilde{P}_1$ and 
$^3 \tilde{P}_1$ states, respectively.
By inserting the initial and the final wave functions into 
the Schr\"{o}dinger equation (\ref{eq:schro}) with the
strong and weak PNC potentials,
one can obtain the radial wave equation for each channel (see Appendix 
for details).

\section{Matrix Elements, $P_\gamma$ and $A_\gamma$}

At the threshold energy, it is well known that the neutron capture 
cross section is dominated by the isovector M1 transition.
We can evaluate the parity conserving M1 transition amplitude
by using the one-body spin current operator
\begin{equation}
\mbfJ_{M} = -i\, \frac{\mu_V}{4 m_N} \sum_{i} \tau^z_i \,
\vsig_i \times \mbfk_\gamma,
\end{equation}
where
$\mu_V = 4.71$, and $\mbfk_\gamma$ is the photon momentum.
Amplitudes between the states with opposite parities would
become non-zero through the E1 transition.
While the impulse approximation is used in evaluating the M1 amplitude,
the contribution from the exchange currents can be well accounted for
by the Siegert's theorem. 
The E1 current operator with Siegert's theorem
reads
\begin{equation}
\mbfJ^S_{E} = -i\, \frac{\omega}{4}\, (\tau^z_1 - \tau^z_2) \mbfr,
\end{equation}
where $\omega$ is the photon energy (2.2246 MeV at threshold).
The transition amplitudes (${\cal M}^f_i$)
can be classified in terms of the 
electromagnetic type ${\cal M}$(=$M$ or $E$), the initial ($i$) and the 
final ($f$) states.
The leading parity-conserving isovector M1 transition occurs between 
the initial $^1 S_0$ and final $^3 S_1$ states, 
and we denote its amplitude by $M^{3s1}_{1s0}$.
The non-zero PNC E1 amplitudes are represented similarly as
$\tilde{E}^{3s1}_{3p0}$ for
$^3 \tilde{P}_0 \rightarrow {}^3 S_1 + {}^3 D_1$, 
$\tilde{E}^{1p1}_{1s0}$ for $^1 S_0 \rightarrow {}^1 \tilde{P}_1$,
$\tilde{E}^{3s1}_{3p1}$ for $^3 \tilde{P}_1 \rightarrow
{}^3 S_1 + {}^3 D_1$, and 
$\tilde{E}^{3p1}_{3s1}$ for $^3 S_1 + {}^3 D_1 \rightarrow
{}^3 \tilde{P}_1$, where the tildes are to distinguish the 
PNC amplitudes from the normal parity conserving ones.
With the wave functions of Eqs.~(\ref{eq:pci})--(\ref{eq:pncf}), 
we obtain the matrix elements
\begin{eqnarray}
M^{3s1}_{1s0} &=& \frac{\omega\, \mu_V}{4 m_N} 
\int dr\, u_d(r)\, u_s(r), \label{eq:M1}\\
\tilde{E}^{3s1}_{3p0} &=& \frac{\omega}{12} \int\, dr\, r 
\left( u_d(r) - \sqrt{2} w_d(r) \right) \tilde{v}^{3p0}_t(r), 
\label{eq:E3p0}\\
\tilde{E}^{1p1}_{1s0} &=& \frac{\omega}{4 \sqrt{3}}
\int dr\, r\, \tilde{v}^{1p1}_d(r)\, u_s(r), 
\label{eq:E1p1}\\
\tilde{E}^{3s1}_{3p1} &=& - \frac{\omega}{4 \sqrt{6}}
\int dr\, r\, \left( u_d(r) + \frac{w_d(r)}{\sqrt{2}} \right)\, 
\tilde{v}^{3p1}_t, 
\label{eq:E3p1s}\\
\tilde{E}^{3p1}_{3s1} &=& \frac{\omega}{4 \sqrt{6}}
\int dr\, r\, \tilde{v}^{3p1}_d\, 
\left( u_t(r) + \frac{w_t(r)}{\sqrt{2}}\right).
\label{eq:E3p1d}
\end{eqnarray}
In terms of these electromagnetic amplitudes, 
the two PNC observables are written as
\begin{eqnarray}
A_\gamma &=& -2\, \frac{\tilde{E}^{3s1}_{3p1} +
\tilde{E}^{3p1}_{3s1}}{M^{3s1}_{1s0}} \equiv A^i_\gamma + A^f_\gamma,
\label{eq:agamma} \\
P_\gamma &=& -2\, \frac{\tilde{E}^{3s1}_{3p0} +
\tilde{E}^{1p1}_{1s0}}{M^{3s1}_{1s0}} \equiv P^i_\gamma + P^f_\gamma, 
\label{eq:pgamma}
\end{eqnarray}
where 
$A^i_\gamma \equiv -2 \tilde{E}^{3s1}_{3p1}/M^{3s1}_{1s0}$ and
$P^i_\gamma \equiv -2 \tilde{E}^{3s1}_{3p0}/M^{3s1}_{1s0}$
have the PNC component in the initial state and
$A^f_\gamma \equiv -2 \tilde{E}^{3p1}_{3s1}/M^{3s1}_{1s0}$ and
$P^f_\gamma \equiv  -2 \tilde{E}^{1p1}_{1s0}/M^{3s1}_{1s0}$
have the PNC component in the final state.
Numerical results are given in Table~\ref{tab:result}.
We express the results for $A_\gamma$ and
$P_\gamma$ in terms of the weak coupling constants $h^{\Delta I}_M$
to show explicitly the dependence of $A_\gamma$ and $P_\gamma$ on each
meson.
``Best values'' refer to $A_\gamma$ and $P_\gamma$ values 
evaluated with the so-called best values of the weak
meson-coupling constants suggested by Ref.~\cite{ddh80}.
They are 
$h^0_\rho = -11.4$,
$h^0_\omega = -1.9$
$h^2_\rho = -9.5$,
$h^1_\pi = 4.6$, 
$h^1_\rho = -0.2$
and 
$h^1_\omega = -1.1$, in units of $10^{-7}$. 
%%%%%%%%%%%%%%%%%%%% Table 1 %%%%%%%%%%%%%%%%%%%%%%%%%%%%%%%%%%%%
\begin{table}[tbp]
\begin{center}
\begin{tabular}{|c|c|c|c|c|}\hline
 &     & \multicolumn{3}{|c|}{Best values ($\times 10^{-8}$)} \\ \cline{3-5}
Model & $A_\gamma$ & $A_\gamma$ & $A^i_\gamma$ & $A^f_\gamma$ \\ \hline
Paris  & $-0.148 h^1_\pi - 0.001 h^1_\rho + 0.003 h^1_\omega$ & 
$-6.85$  & $-3.34$ & $-3.51$ \\ \hline
Bonn   & $-0.117 h^1_\pi - 0.001 h^1_\rho + 0.003 h^1_\omega$ &
$-5.42$  & $-2.66$ & $-2.76$ \\ \hline
Bonn-B & $-0.117 h^1_\pi - 0.001 h^1_\rho + 0.002 h^1_\omega$ &
$-5.41$  & $-2.65$ & $-2.76$  \\ \hline
A$v18$ & $-0.117 h^1_\pi - 0.001 h^1_\rho + 0.002 h^1_\omega$ &
$-5.41$  & $-2.63$ & $-2.78$  \\ \hline\hline
 &     &\multicolumn{3}{|c|}{Best values ($\times 10^{-8}$)} \\ \cline{3-5}
Model & $P_\gamma$ & $P_\gamma$ & $P^i_\gamma$ & $P^f_\gamma$ \\ \hline
Paris  & $-0.0106 h^0_\rho + 0.0074 h^0_\omega - 0.0191 h^2_\rho$ & 
2.88   & $-1.24$ & 4.12 \\ \hline
Bonn   & $-0.0890 h^0_\rho + 0.0088 h^0_\omega - 0.0214 h^2_\rho$ &
12.01  & $-1.40$ & 13.4 \\ \hline
Bonn-B & $-0.0286 h^0_\rho + 0.0012 h^0_\omega - 0.0208 h^2_\rho$ &
5.21   & $-1.35$ & 6.56 \\ \hline
A$v18$ & $-0.0088 h^0_\rho + 0.0034 h^0_\omega - 0.0175 h^2_\rho$ &
2.64   & $-1.11$ & 3.75 \\ \hline
\end{tabular}
\end{center}
\caption{Results for the observables $A_\gamma$ and $P_\gamma$ 
for various phenomenological models in terms of the weak 
coupling constants $h^{\Delta I}_M$.
Best values mean $A_\gamma$ and $P_\gamma$ values obtained
with the best values of $h^{\Delta I}_M$ suggested by DDH.
They are in units of $10^{-8}$.
The definitions of $P^{i,f}_\gamma$ and $A^{i,f}_\gamma$ are given in
Eqs.~(\ref{eq:pgamma}) and (\ref{eq:agamma}), respectively.}
\label{tab:result}
\end{table}
%%%%%%%%%%%%%%%%%%%%%%%% End of Table 1 %%%%%%%%%%%%%%%%%%%%%%%%%%%%
 
\section{Results and Discussions}

\noindent
{\large Asymmetry (${\bm A_\gamma}$)}

\begin{figure}
\begin{center}
\epsfig{file=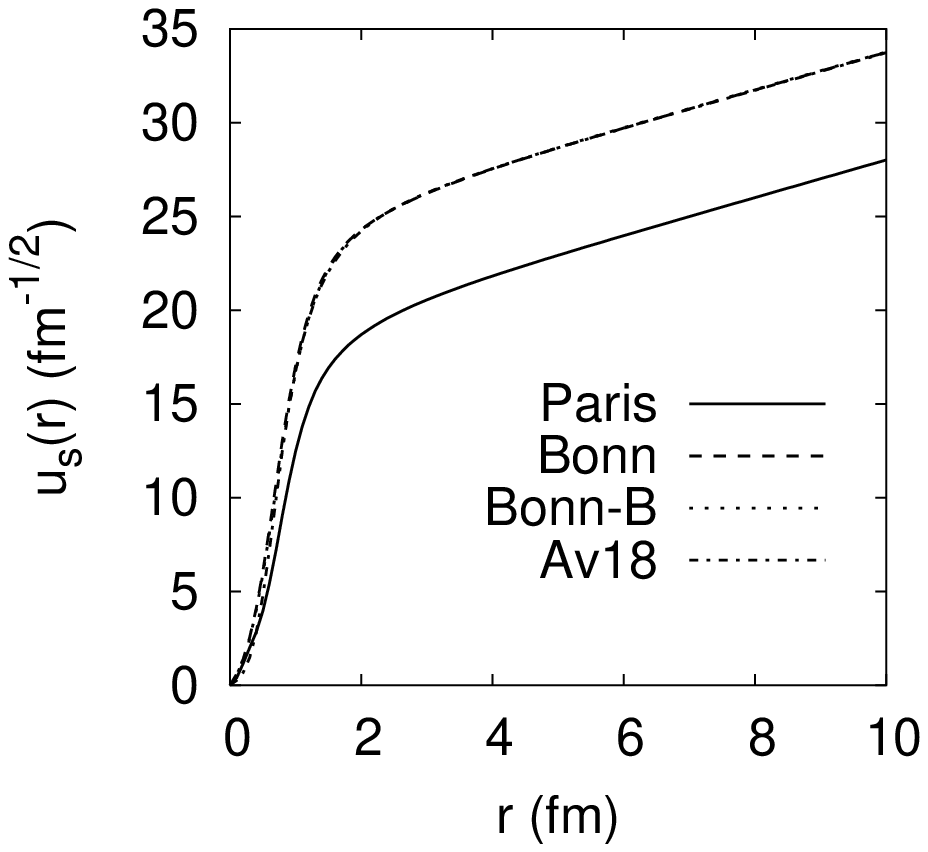, width=5.3cm}
\epsfig{file=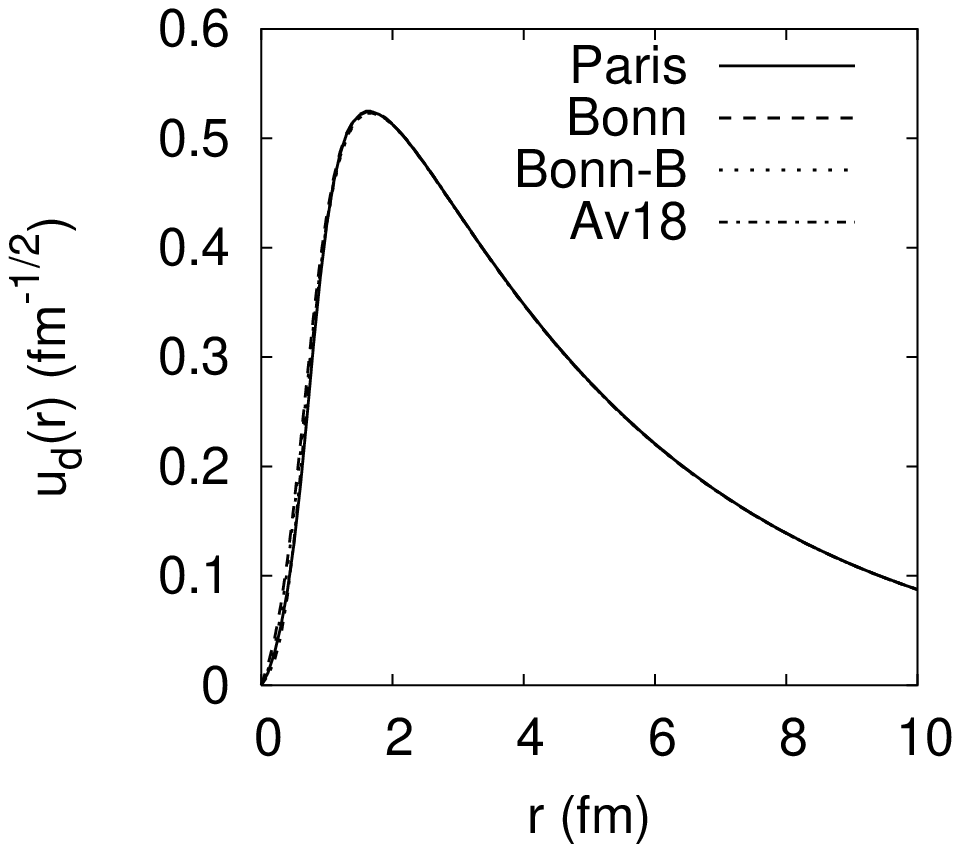, width=5.3cm}
\epsfig{file=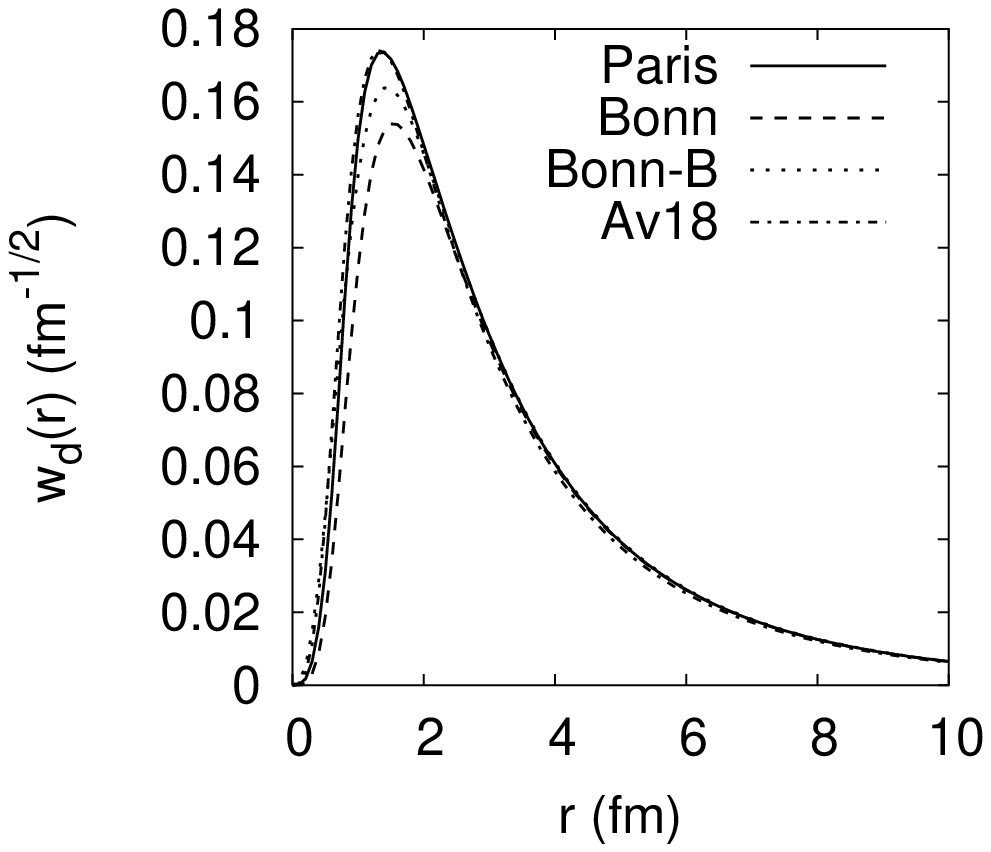, width=5.3cm}
\end{center}
\caption{The wave functions $u_s(r)$, $u_d(r)$ and $w_d(r)$ 
calculated with different potentials are plotted. 
The results for $u_s(r)$ from Bonn, Bonn-B and 
A$v18$ are indistinguishable and 
correspond to the upper curve in the figure.
$u_t(r)$ and $w_t(r)$ are not shown here, because they do not depend
very much on the models.}
\label{fig:pc-wf}
\end{figure}

\begin{figure}
\begin{center}
\epsfig{file=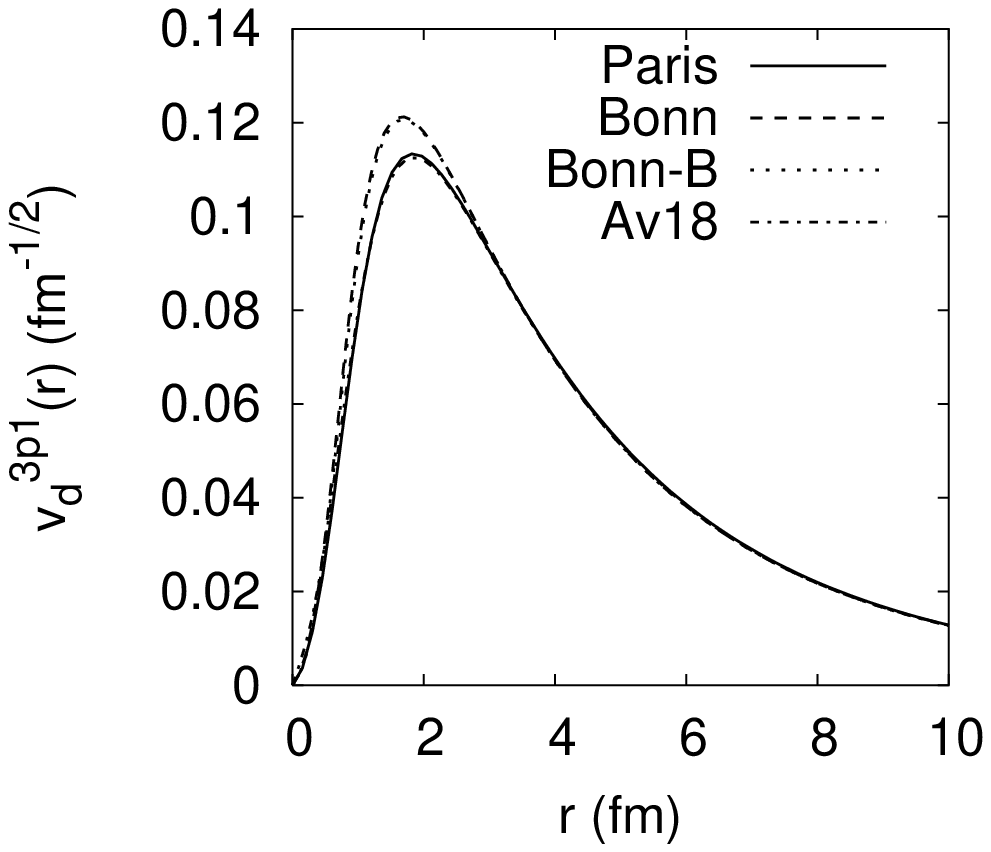, width=5.3cm}
\epsfig{file=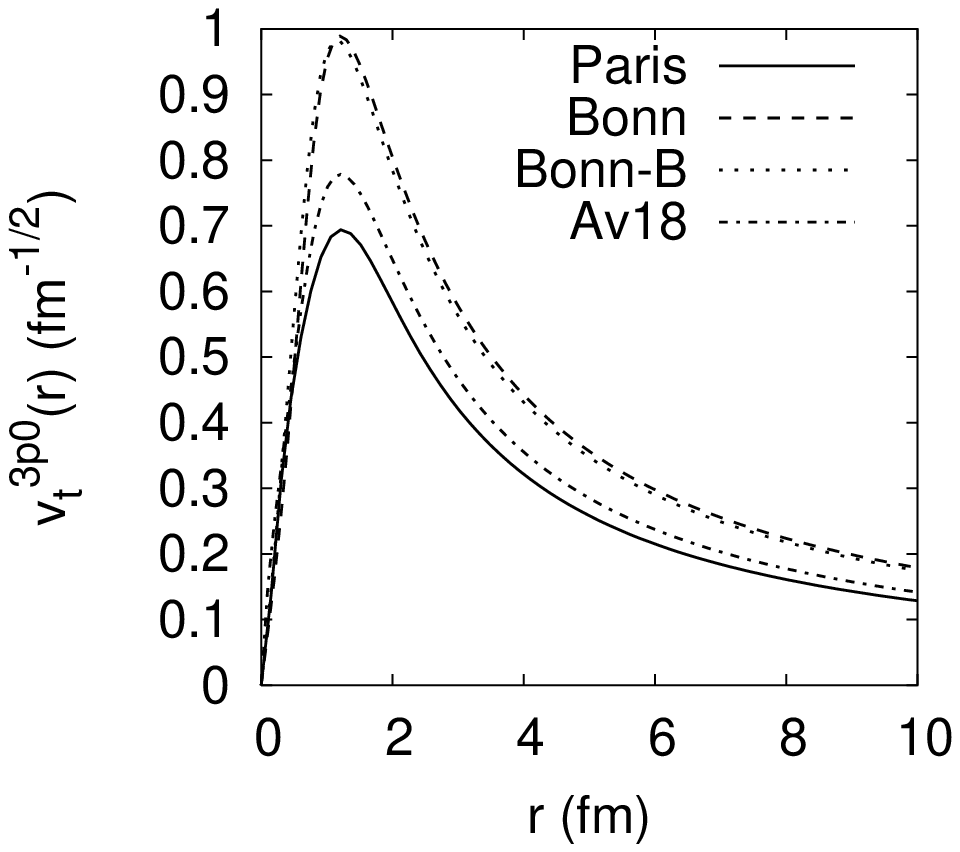, width=5.3cm}
\epsfig{file=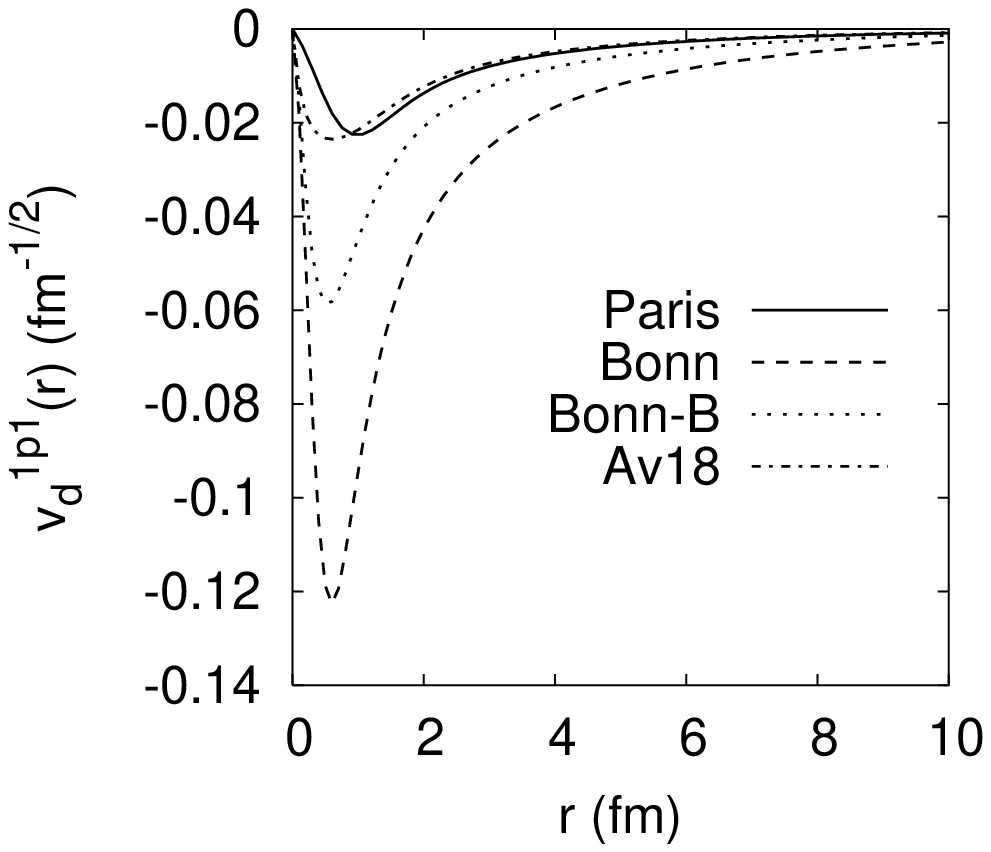, width=5.3cm}
\end{center}
\caption{Wave functions for the parity-admixed states 
for different potentials.
The wave functions are given in units of $h^1_\pi$.
Note the difference in the scale.}
\label{fig:pnc-wf}
\end{figure}

As shown in Table~\ref{tab:result}, the
Bonn and A$v18$ models predict the same $A_\gamma$ value,
while the best value from the Paris potential is larger in magnitude
than those from Bonn and A$v18$ by a factor of 1.27.
This factor can be understood by examining the wave functions
that contribute to $A_\gamma$. $u_s$ and $u_d$ are plotted in 
Fig.~\ref{fig:pc-wf}, and $\tilde{v}^{3p1}_d$ in Fig.~\ref{fig:pnc-wf}.
%
%%%%%%%%%%%%%%%%% Figure 1 %%%%%%%%%%%%%%%%%%%%%%%%%%%%%%%%%%%%%%%%%%
%%%%%%%%%%%%%%%% Figure 2 %%%%%%%%%%%%%%%%%%%%%%%%%%%%%%%%%%%%%%%%%%
%%%%%%%%%%%%%%%%%%%%%%%%%%%%%%%%%%%%%%%%%%%%%%%%%%%%%%%%%%%%%%%%%%%%
%
As can be seen in Fig.~\ref{fig:pc-wf} and \ref{fig:pnc-wf},
$u_d$ and $\tilde{v}^{3p1}_d$ calculated with different
potentials are very similar to each other for all potentials,
but $u_s$ calculated with the Paris potential is substantially different
from $u_s$ from other potentials. ($\tilde{v}^{3p1}_t$, though not shown here,
is more or less the same for all potential models.)
The reason for this difference can be traced back to the fact that
the Paris potential was fitted to proton-proton data and therefore
yields a scattering length of $a = -17.6$ fm 
for the $^1 S_0$ channel (in the absence of the Coulomb interaction)  
while A$v18$ and Bonn are neutron-proton models and yield 
$a = -23.7$ fm.
The M1 isovector amplitude $M^{3s1}_{1s0}$ in the denominator
of Eqs.~(\ref{eq:agamma}) and (\ref{eq:pgamma})
is 0.184 for Paris potential and 0.233 for A$v18$ and Bonn potentials. 
Since the wave functions that contribute to the E1 amplitude 
in Eqs.~(\ref{eq:E3p1s}) and (\ref{eq:E3p1d})
(numerator of $A_\gamma$) are very similar, 
the difference in the values of $A_\gamma$ comes mostly from the value of the 
M1 amplitude in the denominator of $A_\gamma$.
Indeed, the ratio of the M1 amplitudes for A$v18$ to Paris, 
1.27, is equal to the ratio of the best value of $A_\gamma$.
Thus, if one could readjust the Paris potential to produce 
the accepted $n-p$ scattering length in the $^1 S_0$ channel, the
four models would give us essentially model-independent $A_\gamma$ values.
A recent work \cite{scp-04} in which various contributions from the
exchange currents are taken into account confirms this model-independent
nature of $A_\gamma$.
The magnitude of $A_\gamma$ in Ref. \cite{scp-04} 
($A_\gamma = - 4.98 \times 10^{-8}$)
with pion-exchange currents 
is smaller than ours ($A_\gamma = - 5.41 \times 10^{-8}$) 
by about 9 \%, 
and is in agreement with the result of Ref. \cite{hyun01} 
($A_\gamma = -4.94 \times 10^{-8}$)
where one-body and leading pion-exchange currents are considered.
On the other hand, if one employs the $h^1_\pi$ value from the
$^{18}$F \cite{ah85} and $^{133}$Cs \cite{cs133} experiments, 
$A_\gamma$ becomes
$-1.52 \times 10^{-8}$ and $-11.1 \times 10^{-8}$, respectively.
Since the contribution from the pion to $A_\gamma$ is 
more than 99 \% of the total value (see Table~\ref{tab:result}), 
a more accurate measurement of $A_\gamma$ can provide a stringent 
determination of $h^1_\pi$.

In passing, we remark that $A_\gamma$ from a previous work \cite{mga-npa86}
using Paris potential differs from our $A_\gamma$ in sign though 
the magnitudes agree.
It appears that the definitions of $A_\gamma$ differ in sign.

\noindent
{\large Polarization (${\bm P_\gamma}$)}

While $A_\gamma$ is dominated by the long range part of the interactions
and is practically model-independent,
$P_\gamma$ depends strongly on the heavy meson exchanges and 
on the potential model.
$P_\gamma$'s calculated with the best values of the weak coupling 
constants \cite{ddh80} and Paris and A$v18$ potentials 
are similar to each other,
but $P_\gamma$'s evaluated with Bonn and Bonn-B are larger than 
that with A$v18$ by a factor of 5 and 2,
respectively. (Bonn-A and Bonn produce similar results and thus
Bonn-A is not included in the discussion.)
$P_\gamma$'s expressed in terms of $h^{\Delta I}_M$ in Table~\ref{tab:result}
show that $P_\gamma$ from Bonn is more sensitive to $h^0_\rho$ than
$P_\gamma$ from other potentials, while the terms depending on $h^0_\omega$ 
and $h^2_\rho$ are only moderately model dependent.
The contributions from the 
initial ($^3\tilde{P}_0$) and the final ($^1\tilde{P}_1$) states
listed in Table~\ref{tab:result} show that the initial state 
contribution ($P^i_\gamma$) is rather model-independent, 
but the contribution from the final state ($P^f_\gamma$)
is highly dependent on the potentials.

\begin{figure}[tbp]
\begin{center}
\epsfig{file=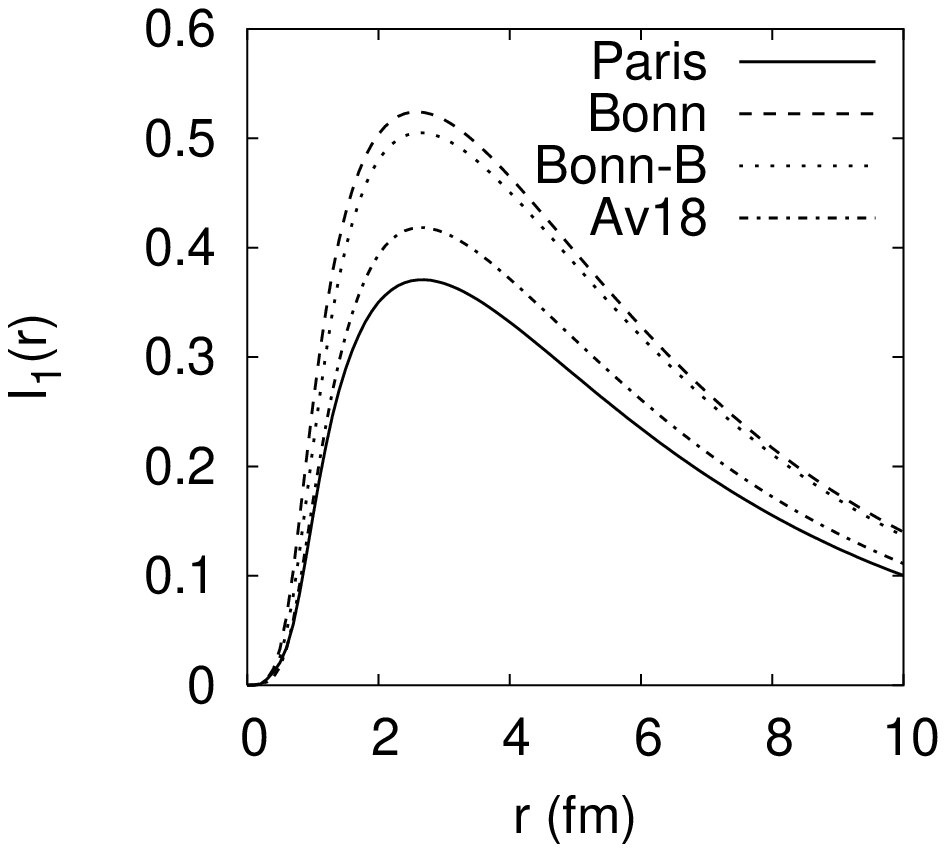, width=6.0cm}
\epsfig{file=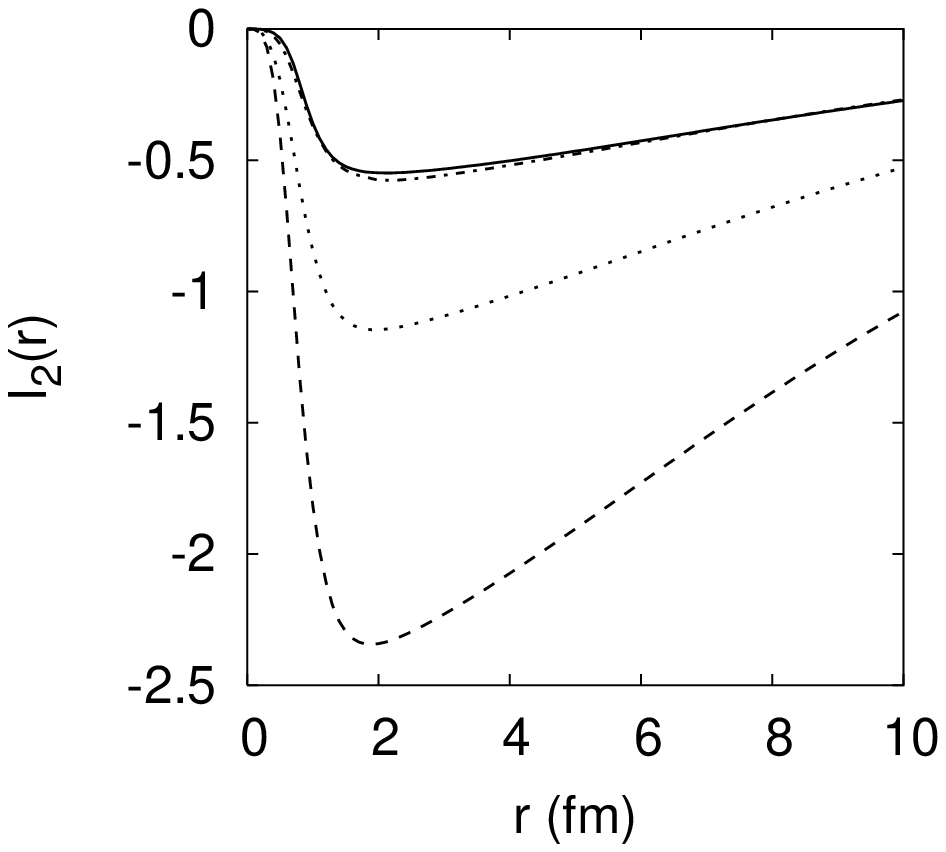, width=6.0cm}
\end{center}
\caption{Integrands that enter into the E1 amplitude of $P_\gamma$.
The left panel shows 
$I_1(r) \equiv r\, (u_d(r) - \sqrt{2} w_d(r))\,\tilde{v}^{3p0}_t(r)$
of Eq.~(\ref{eq:E3p0}), and the right one $I_2(r) \equiv 
r\,\tilde{v}^{1p1}_d(r)\, u_s(r)$ of Eq.~(\ref{eq:E1p1}).}
\label{fig:pnc-int}
\end{figure}

\begin{figure}[tbp]
\begin{center}
\epsfig{file = 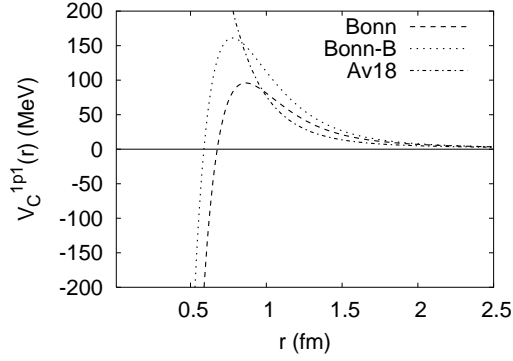, width=7cm}
\end{center}
\caption{Comparison of the central potentials in the $^1 P_1$ channel  
of the Bonn, Bonn-B, and A$v18$ potentials.}
\label{fig:pot1p1}
\end{figure}

The numerical factors in front of the weak coupling constants 
in Table~\ref{tab:result} are determined by
the strong potentials in each channel through the wave functions of
the $^1 S_0$ and $^3 S_1 - {}^3 D_1$ channels that enter
into the source terms in the Schr\"{o}dinger equation of the parity-admixed
states (see Appendix). 
The wave functions for the $^3\tilde{P}_0$ and the $^1\tilde{P}_1$ states
are shown in Fig.~\ref{fig:pnc-wf}, and the corresponding integrands
of the E1 amplitudes, Eqs.~(\ref{eq:E3p0}) and (\ref{eq:E1p1}), are 
shown in Fig.~\ref{fig:pnc-int}.
%%%%%%%%%%%%%%%%%%%%%%%%%%%%%%%% Fig. 3 %%%%%%%%%%%%%%%%%%%%%%%%%%%%%%%%%
%%%%%%%%%%%%%%%%%%%%%%%%%%%%%%%%%%%%%%%%%%%%%%%%%%%%%%%%%%%%%%%%%%%%%%%%
The wave function ($\tilde{v}^{3p0}_t(r)$)
and the integrand ($I_1(r)$) for the $^3\tilde{P}_0$ state exhibit
a sizeable model dependence. However, for the $^1\tilde{P}_1$ channel there
are even more drastic variations in both wave function 
($\tilde{v}^{1p1}_d (r)$)
and the integrand ($I_2(r)$) depending on different potentials. 
Such a strong model dependency can be understood
from the behavior of the strong potential in the $^1 P_1$ channel 
and the source term that contributes to $\tilde{v}^{1p1}_d$ in the 
Schr\"{o}dinger equation (see Eq. (\ref{eq:eq41p1})).
Fig.~\ref{fig:pot1p1} shows the strong potentials in the $^1 P_1$ channel.
%%%%%%%%%%%%%%%%%% Fig. 4 %%%%%%%%%%%%%%%%%%%%%%%%%%%%%%%%%%%%%%%%
%%%%%%%%%%%%%%%%%%%%%%%%%%%%%%%%%%%%%%%%%%%%%%%%%%%%%%%%%%%%%%%%%%%
The Bonn potential for $^1 P_1$ channel becomes attractive in the short
range region while A$v18$ is repulsive in the whole region.
The attraction at short ranges increases the probability for a nucleon
to be present in the region, 
and this can partly explain the shape of $\tilde{v}^{1p1}_d$
in Fig.~\ref{fig:pnc-wf}. 
A recent work by R. Schiavilla {\it et al.}
shows a similar behavior of $\tilde{v}^{1p1}_d$ \cite{scp-04}.

%%%%%%%%%%%%%%%%%%%% Fig. 5 %%%%%%%%%%%%%%%%%%%%%%%%%%%%%%%%%%%%%%%
\begin{figure}[tbp]
\begin{center}
\epsfig{file= 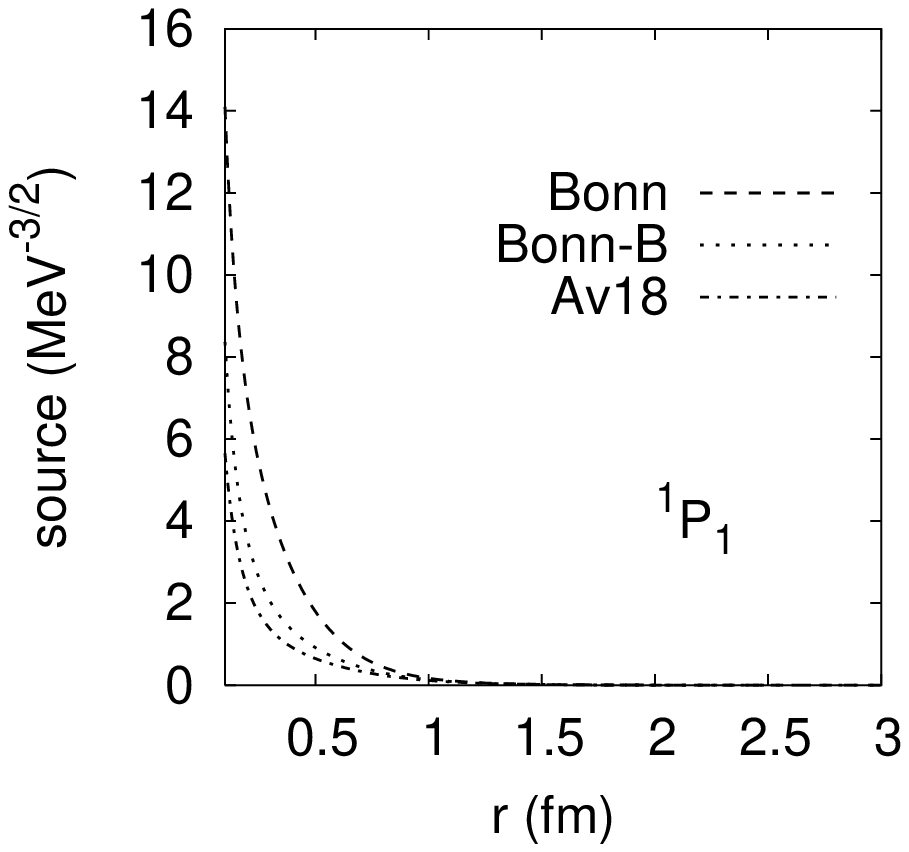, width = 7cm}
\epsfig{file= 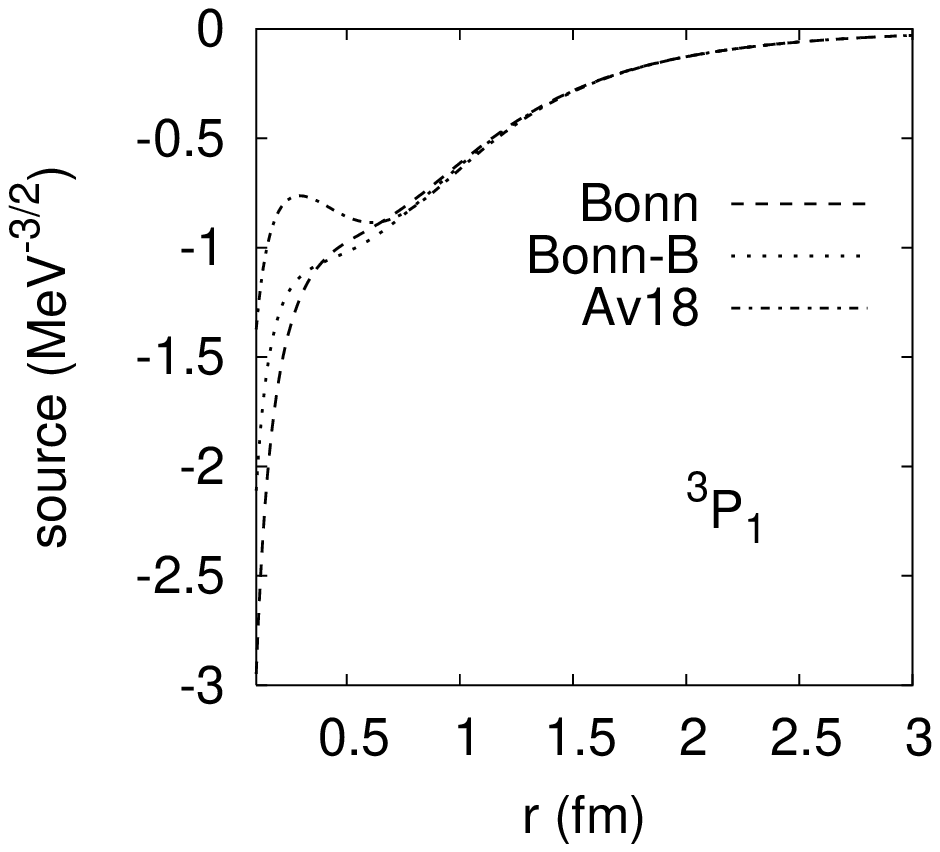, width = 7cm}
\end{center}
\caption{Source terms for the $^1\tilde{P}_1$ and 
$^3\tilde{P}_1$ states. Compared to the terms of $^3 P_1$,
those of $^1 P_1$ are more model dependent. 
Note the difference in the scale.}
\label{fig:source}
\end{figure}
%%%%%%%%%%%%%%%%%%%%%%%%%%%%%%%%%%%%%%%%%%%%%%%%%%%%%%%%%%%%%%%%%%%

In Fig.~\ref{fig:source}, we compare the source terms of the 
$^3\tilde{P}_1$ (the right hand side of Eq.~(\ref{eq:eq43p1}))
and $^1\tilde{P}_1$ (the right hand side of Eq.~(\ref{eq:eq41p1}))
states.
The sources for the $^3\tilde{P}_1$ state from different models
show a moderate model dependence
for $r \leq 0.5$ fm, but they have significant magnitudes and 
are indistinguishable in the 
intermediate and long range regions, which explains the 
model independent results of $A_\gamma$ in Table~\ref{tab:result}.
On the contrary, most of the contribution to the source terms of
$\tilde{v}^{1p1}_d$ comes from the intermediate and short range region,
and they depend strongly on the model. 
A larger source combined with attraction in the short-range region, 
as it is the case for the $^1\tilde{P}_1$ channel of the Bonn models,
yields an enhanced contribution to $P_\gamma$.

Concluding, we have calculated parity non-conserving 
observables $P_\gamma$ for the reaction $n + p \rightarrow d + \gamma$ and 
$A_\gamma$ for the reaction $\vec{n} + p \rightarrow d + \gamma$ at
threshold. We have employed the Paris, Bonn and A$v18$ potentials
for the strong interaction and the DDH potential for the weak interaction. 
$A_\gamma$ turns out to be independent of the strong interaction models,
while $P_\gamma$ is sensitive to the dynamics at short ranges. 
Since $A_\gamma$ is rather strong-interaction independent,
one can reduce the uncertainty in the value of $h^1_\pi$ 
by measuring $A_\gamma$ accurately.
Regarding $P_\gamma$, there are relatively large uncertainties, which 
stem from ambiguities in both strong and weak interactions at 
short ranges. 
However, since the major uncertainty comes from the
$^1 \tilde{P}_1$ channel and the value of $h^0_\rho$, 
an accurate experimental measurement of $P_\gamma$ can shed 
some light on the weak coupling constants.

\section*{Acknowledgments}

We thank B. Desplanques for useful discussions.
The work is supported by the Korea Research 
Foundation Grant (KRF-2003-070-C00015).

\appendix \label{sec:appen}
\setcounter{equation}{0}
\renewcommand{\theequation}{A.\arabic{equation}}
\section*{Appendix}

%In this section, we summarize the Schr\"{o}dinger equations for the 
%wave functions that contribute to $P_\gamma$ and $A_\gamma$.
The radial equations for the $^1 S_0$ continuum, and the $^3 S_1 - {}^3 D_1$
continuum and bound states read 
\begin{eqnarray}
u''_s(r) + m_N\, \left(E - V_C(r) \right)\, u_s(r) &=& 0, \\
u''_{t(d)}(r) + m_N\, \left(E - V_C(r)\right) u_{t(d)}(r) 
&=& \sqrt{8} m_N V_T(r) w_{t(d)}(r), \\
w''_{t(d)}(r) - \frac{6}{r^2}w_{t(d)}(r) 
- m_N\, (E - V_C(r) + 2 V_T(r)) w_{t(d)}(r)
&=& \sqrt{8} m_N V_T(r) u_{t(d)}(r).
\end{eqnarray}
The equations for the parity-admixed states are
\begin{eqnarray}
%%%%%%%%%%%%%%%% equation for 3p0 %%%%%%%%%%%%%%%%%%%%%%%%%%%%%%%%%%%%
& &\tilde{v}^{3p0 ''}_t(r) - \frac{2}{r^2}\, \tilde{v}^{3p0}_t
+ m_N\, (E - V_C(r) + 4 V_T(r) ) \, \tilde{v}^{3p0}_t = \nonumber \\
& & -2 \left[(\chi_\rho + 2)\, u_s(r) \frac{\partial}{\partial r}
\left( F^0_\rho(r) - \sqrt{\frac{2}{3}} F^2_\rho(r) \right)
+ (\chi_\omega + 2)\, u_s(r)\, \frac{\partial}{\partial r} F^0_\omega(r)
\right. \nonumber \\ & & \left.
+ 2 r\, \left( F^0_\rho(r) - \sqrt{\frac{2}{3}} F^2_\rho(r)
+ F^0_\omega(r) \right)
\frac{\partial}{\partial r} \left( \frac{u_s(r)}{r} \right) \right], \\
%\equiv \tilde{J}^{3p0}_t, \label{eq:eq43p0} \\
%%%%%%%%%%%%%%%% equation for 3p1 %%%%%%%%%%%%%%%%%%%%%%%%%%%%%%%%%%%%
& &\tilde{v}^{3p1 ''}_{t(d)}(r) - \frac{2}{r^2}\, \vto + m_N\,
(E - V_C(r) - 2 V_T(r))\, \vto = \nonumber \\
& &\frac{2}{\sqrt{3}} \left[
\left( u_{t(d)}(r) + \frac{1}{\sqrt{2}} w_{t(d)}(r) \right)
\frac{\partial}{\partial r} \left(
F^1_\pi(r) + \sqrt{2} F^1_\rho(r) - \sqrt{2} F^1_\omega(r) 
- \sqrt{2} F^{1'}_\rho(r) \right) \right.
\nonumber \\
& & \hspace{1cm} + 2 \sqrt{2} \left(F^1_\rho(r) - F^1_\omega(r)\right)
\frac{\partial}{\partial r}
\left( u_{t(d)}(r) + \frac{1}{\sqrt{2}} w_{t(d)}(r) \right)
\nonumber \\
& &  \hspace{1cm} \left. - \frac{2 \sqrt{2}}{r}
\left( F^1_\rho(r) - F^1_\omega(r)\right) 
\left(u_{t(d)}(r) - \sqrt{2} w_{t(d)}(r)\right)
\right], \label{eq:eq43p1} \\
% \equiv \tilde{J}^{3p1}_{t(d)},\\ 
%%%%%%%%%%%%%%%%% equation for 1p1 %%%%%%%%%%%%%%%%%%%%%%%%%%%%%%%%%%%%
& &\tilde{v}^{1p1 ''}_{d}(r) - \frac{2}{r^2} \tilde{v}^{1p1}_d + m_N\, 
(E - V_C(r) )\, \tilde{v}^{1p1}_d = \nonumber \\
& &\frac{2}{\sqrt{3}} \left[ \left(u_{d}(r) - \sqrt{2} w_{d}(r)\right)
\frac{\partial}{\partial r} \left(
3 \chi_\rho F^0_\rho(r) - \chi_\omega F^0_\omega(r) \right) \right.
\nonumber \\
& &\hspace{1cm} - 2 \left(3 F^0_\rho(r) - F^0_\omega(r)\right)
\frac{\partial}{\partial r} \left(u_{d}(r) - \sqrt{2} w_{d}(r)\right)
\nonumber \\
& & \hspace{1cm} \left. + \frac{2}{r} 
\left(3 F^0_\rho(r) - F^0_\omega(r)\right)
\left(u_{d}(r) +2\sqrt{2} w_{d}(r)\right)\right],
% \equiv \tilde{J}^{1p1}_d
\label{eq:eq41p1}
\end{eqnarray}
where
$F^{\Delta I}_M(r) \equiv g_{M NN}\, h^{\Delta I}_M\, f_M(r)$.
%and $F^{1'}_\rho(r) \equiv g_{\rho NN}\, h^{1'}_\rho\, f_\rho(r)$.
%%%%%%%%%%%%%%%%%%%%%%%%%%%%%%%%%%%%%%%%%%%%%%%%%%%%%%%%%%%%%%%%%%%%%%%%%%5


\begin{thebibliography}{99}
\bibitem{cs133} C. S. Wood {\it et al.}, Science {\bf 275} (1997) 1759.
\bibitem{bwprl-99} S. C. Bennett and C. E. Wieman, Phys. Rev. Lett.
{\bf 82} (1999) 2484.
\bibitem{pp} A. R. Berdoz {\it et al.}, Phys. Rev. Lett. 
{\bf 87} (2001) 272301.
\bibitem{np} W. M. Snow {\it et al.}, Nucl. Instrum. Methods
{\bf 440} (2000) 729.
\bibitem{ddh80} B. Desplanques, J. F. Donoghue and B. R. Holstein,
Ann. Phys. {\bf 124} (1980) 449.
\bibitem{ah85} E. G. Adelberger and W. C. Haxton, 
Ann. Rev. Nucl. Part. Sci. {\bf 35} (1985) 501.
\bibitem{des-physrep} B. Desplanques, Phys. Rep. {\bf 297} (1998) 1.
\bibitem{ill88} J. Alberi {\it et al.}, Can. J. Phys.
{\bf 66} (1988) 542.
\bibitem{des75} B. Desplanques, Nucl. Phys. {\bf A 242} (1975) 425.
\bibitem{paris} M. Lacombe, B. Loiseau, J. M. Richard,
R. Vinh Mau, J. C\^{o}t\'{e}, P. Pir\'{e}s and R. de Tourreil,
Phys. Rev. C {\bf 21} (1980) 861.
\bibitem{bonn87} R. Machleidt, K. Holinde and Ch. Elster,
Phys. Rept. {\bf 149} (1987) 1.
\bibitem{bonn-ab} R. Machleidt, Adv. Nucl. Phys. {\bf 19}
(1989) 189.
\bibitem{av18} R. B. Wiringa, V. G. J. Stoks and
R. Schiavilla, Phys. Rev. C {\bf 51} (1995) 38.
\bibitem{cra76} B. A. Craver, E. Fischbach, Y. E. Kim and A. Tubis,
Phys. Rev. D {\bf 13} (1976) 1376.
\bibitem{pgam} V. A. Knyazkov {\it et al.}, 
Nucl. Phys. {\bf A 417} (1984) 209.
\bibitem{mck78} B. H. H. McKellar and K. R. Lassey,
Phys. Rev. C {\bf 17} (1978)842.
\bibitem{pmetalnpa81} F. Pauss, L. Mathelitsch, J. C\^{o}t\'{e},
M. Lacombe, B. Loiseau and R. Vinh Mau, Nucl. Phys. {\bf A 365} (1981)
392.
\bibitem{scp-04} R. Schiavilla, J. Carlson and M. Paris, 
Phys. Rev. C {\bf 70} (2004) 044007.
\bibitem{hyun01} C. H. Hyun, T.-S. Park and D.-P. Min,
Phys. Lett. {\bf B 516} (2001) 321.
\bibitem{mga-npa86} S. Morioka, P. Grange and Y. Avishai,
Nucl. Phys. {\bf A 457} (1986) 518.
%\bibitem{ber01} B. Desplanques, Phys. Lett. {\bf B 512} (2001) 305.
%\bibitem{savage01} M. J. Savage, Nucl. Phys. {\bf A 695} (2001) 365.
%\bibitem{mac-prc01} R. Machleidt, Phys. Rev. C {\bf 63} (2001) 024001.
\end{thebibliography}
\end{document}